\documentclass[twocolumn,aps,letterpaper]{revtex4}
\usepackage{graphicx}
\usepackage{amsmath}
\usepackage{amssymb}
\usepackage{wasysym}

\begin{document}
\title{Experimental characterization of vibrated granular rings}
\author{Z.~A.~Daya$^{1,2}$,  E.~Ben-Naim$^{1,3}$ \&
R.~E.~Ecke$^{1,2,3}$}
\address{${^1}$Center for Nonlinear Studies, ${^2}$Condensed Matter \&
Thermal Physics Group, ${^3}$Theoretical Division, Los Alamos National
Laboratory, Los Alamos, NM 87545}
\date{\today}
\begin{abstract}
We report an experimental study of the statistical properties of vibrated
granular rings. In this system, a linked rod and bead metallic chain in the
form of a ring is collisionally excited by a vertically oscillating plate. The
dynamics are driven primarily by inelastic bead-plate collisions and
are simultaneously constrained by the rings' physical connectedness. By imaging
many instances of the ring configurations, we measure the ensemble averages and
distributions of several physical characteristics on the scale of individual
beads and composite ring.  We study local properties such as inter-bead
seperation and inter-bonds angles, and global properties such as the
radius of gyration and center of mass motion. We characterize scaling
with respect to the size of the chain.

\vspace{0.2cm} \noindent
\end{abstract}
\maketitle
\section{Introduction}
\label{intro}
In the last decade the study of granular materials has become a mainstream
pursuit amongst physicists who have summarized the advances in recent reviews
\cite{kadanoff_99,degennes_99,jaeger_96}. The broad spectrum of static and
dynamic behaviors of granular media are being studied by a combination of
state-of-the-art experiments, computer simulations and theoretical approaches.
Shaken granular materials are one example of a comprehensively studied subset of
this spectrum. Experiments and simulations on shaken grains have demonstrated a
variety of phenomena such as localized structures \cite{umbanhower_96},
extensive patterns \cite{debruyn_98,debruyn_01}, clustering
\cite{kudrolli_97,olafsen_98}, compaction \cite{knight_95,nowak_98} and
non-Maxwellian statistics \cite{olafsen_99,rouyer_00}.

Granular chains consist of hollow spheres connected by rods, {\em i.e.}, grains
constrained by a chain backbone. In granular chains we get the juxtaposition of
the hard grains of granular physics with the flexibility of polymers. The chains
can be excited, say, by collisions with a vibrating plate and can be observed
visually. As such they afford a simple system for the direct measurements of
micro- and macro-scopic variables. Experiments on granular chains, supported
by theoretical models and simulations, have probed the diffusive relaxation of
a simply knotted chain \cite{ebn_01}, the dynamical behavior of a
vertically excited hanging chain \cite{belmonte_01}, the entropic tightening of
a once-twisted ``figure-8'' ring \cite{hastings_02} and the spontaneous
formation of spirals \cite{ecke_03}.

\begin{figure}
\includegraphics[height=9cm]{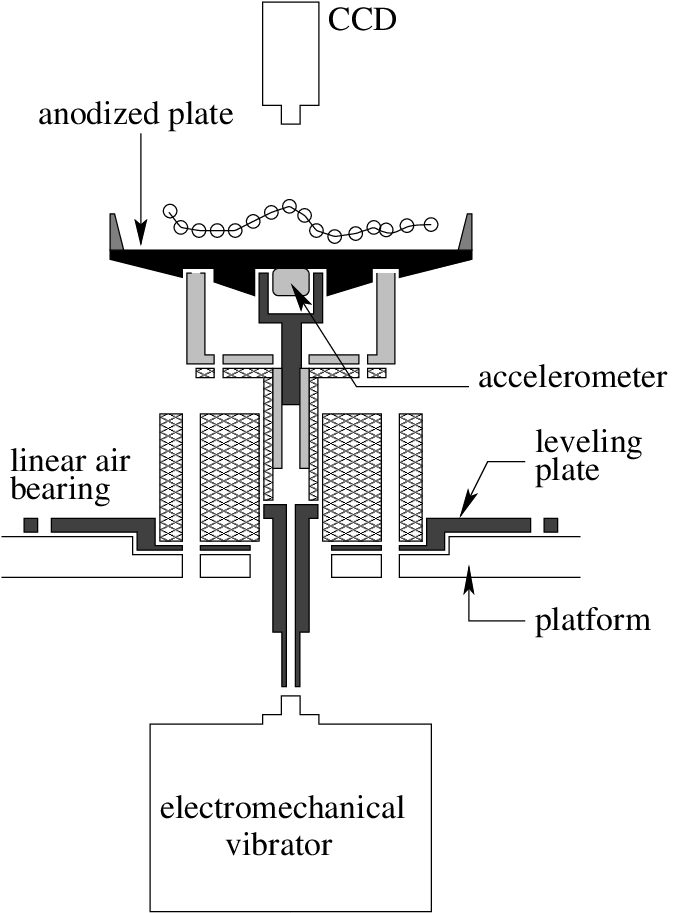}
\caption{\label{expt_fig}
A schematic of our experimental apparatus.}
\end{figure}

In this paper we report our study of vibrated granular rings. Our
experimental system consists of a vertically oscillated plate on which we place
a granular ring constructed from a ball chain; a schematic is shown
in Fig.~\ref{expt_fig}. Rings are constructed by joining the two free ends of
a chain. A chain with free ends suffers unequal excitation along its length
since beads near the ends are less constrained and thus
preferentially excited. Rings, which are devoid of free ends, restore
the equality in excitation amongst the beads, and it is for this reason that
they are particularly attractive to study.

In our experiments we control the amplitude and frequency of the harmonically
oscillated plate. A chain initially resting on the plate becomes airborne when
the plate imparts to it an acceleration that exceeds the gravitational
acceleration. Thereafter, collisions between the plate and chain appear to
randomize the motion and shape of the granular ring. The collisions of different
parts of the chain are asynchronous with the phase of the plate's oscillation.
Further collisions amongst the beads on the chain occur frequently.

We observe the ring from above at a random phase of the oscillation. We acquire
images which are two-dimensional projections of the position of the ring. From
these images we obtain the spatial coordinates and order of the beads in the
ring. Since the images are sequential and time stamped, we are also able to
calculate differences and therefore dynamics. We make a large number of position
measurements for rings under vertical vibration. From this data we study the
geometrical properties on the bead scale, {\em i.e.}, local characteristics, and
on the scale of the chain, {\em i.e.}, global characteristics. The local
properties include the inter-bead separations, the bending angle between two
consecutive rods, the average curvature and the dynamics of the beads. For the
global characteristics we focus on descriptors of the size and overall motion of
the ring. These are the radius of gyration and the dynamics of the rings' center
of mass.

This paper is organized as follows. In Section~\ref{expt} we describe the
experimental apparatus, the specific design considerations, the image
acquisition and analysis procedure. Our experimental results depend at most
on the ordered positions of the beads in a ring and on the data acquisition
rate. We have divided the results into local and global properties which
are discussed in Sections~\ref{local_properties} and \ref{global_properties}
respectively.  Local (global) properties are relevant to the scale of the bead
(ring). We discuss the the inter-bead separation, the inter-rod angle, the
average curvature and the bead dynamics as local properties. For global
characteristics we discuss the radius of gyration and the average motion of
the ring center of mass. A conclusion and an outline of future directions
follows in Section~\ref{conclusion}.

\section{Experiment}
\label{expt}

In this Section we describe our experimental apparatus and our data
extraction procedure. In overall design, the apparatus is comprised of
an electromechanical shaker that drives a plate which is observed by a
CCD camera. A schematic of our experimental setup is shown in
Fig.~\ref{expt_fig}.

\subsection{The electromechanical vibration assembly}
\label{expt:eva}

A commercial electromechanical shaker driven by a sinusoidal voltage input
is used to oscillate a plate. The shaker operates electromagnetically with
the current-carrying coil armature oscillating in the strong magnetic field
of permanent ceramic magnets. The electromechanical shaker is coupled to
the plate by a shaft but is otherwise vibrationally isolated from the rest of
the apparatus. A massive aluminum platform that is rigidly bolted to the
laboratory floor acts as a low-pass filter of vibrational noise. The shaft
is guided by the bore of a precise square-section linear air bearing. In
addition to linear motion, the square-section shaft decouples the plate from the
rotational torque in the drive. The driving is thus accurately uni-axial and a
leveling plate facilitates alignment with gravity. The shaft is coupled to the
plate by two concentric cylindrical tubes. A sensitive accelerometer is attached
centrally on the underside of the plate. 

The plate is driven harmonically with amplitude $A$ and angular velocity $\omega
= 2 \pi f$, where $f$ is the driving frequency. The dimensionless peak
acceleration $\Gamma = A\omega^2/g$, where $g$ is the gravitational
acceleration, and the frequency $f$ specify the driving conditions for the
experiment. The shaker is operated at $f = 16$~Hz and $ \Gamma =2.12$ for all
the experiments described in this paper. The shaker is protected from bearing
the combined weight of the plate and shaft by adjustable counter-weighting
springs such that the drive oscillates about its natural equilibrium position.

\subsection{The plate and the granular rings}
\label{expt:pgr}

The plate is circular with a diameter of $11.50$~inches and is made
of stress-relieved aluminum. The primary design requirements are light
weight, flexural stiffness and surface hardness. Aluminum's low density
and comparable flexural properties to most metals make it a suitable choice.
On oscillation we expect that bending modes of the plate are excited. The higher
the stiffness the higher the frequecies excited. Since our experiments have
drive frequency $f \sim 10$ Hz, we designed the plate so
that its excitable bending modes had much higher frequencies. This is
accomplished by tapering the lower surface of the plate and by boring out
cylinders from this surface in a space-filling azimuthally symmetric manner. The
cylinders bored out of the aluminum have various depths and diameters. The
plate is at its thickest $1.50$ inches decreasing to $0.25$ inches at
the boundary. These considerations lead to a selective damping of pure modes of
the shaken plate. A further safeguard against exciting modes of the plate is
achieved by forcing the plate at two annular regions rather than at the center
or at the edge. The annular regions are at radial locations $1.25 \leq r \leq
2.00$ and $4.88 \leq r \leq 6.00$ inches. Surface hardness needed to withstand
the many metal on metal collisions of the chains hitting the plate is attained
by hard black anodization of the aluminum plate.

The granular chains are commercially available ball chains. They consist of
hollow, approximately-spherical shells that are connected by dumb-bell shaped
rods. We use two types of ball chains in our experiments. The smaller is made
from nickel-plated stainless steel rods and beads. The beads have a nominal
diameter of $3\over32$ inches. The larger is a $1\over8$ inch diameter brass
chain. Thus there are typically $100$ beads across the plate diameter. The
chains are connected end to end to make rings. The splicing together of the
free ends is natural in the sense that a rod from one of the end beads is
inserted into the shell of the other end bead. The joined free ends could seldom
be identified in the constructed ring. The tightest rings that we could make
have
$8$ beads. Hence the angle $\theta$ between two consecutive rods is constrained
so that $0 \leq \theta \leq 50^\circ$. Our experiments use rings with $N$ beads
where $13 \leq N \leq 309$. The rings are laterally confined to the plate by an
acryllic boundary.

\subsection{Data extraction \label{data_extraction}}

\begin{figure}
\includegraphics[height=7.5cm]{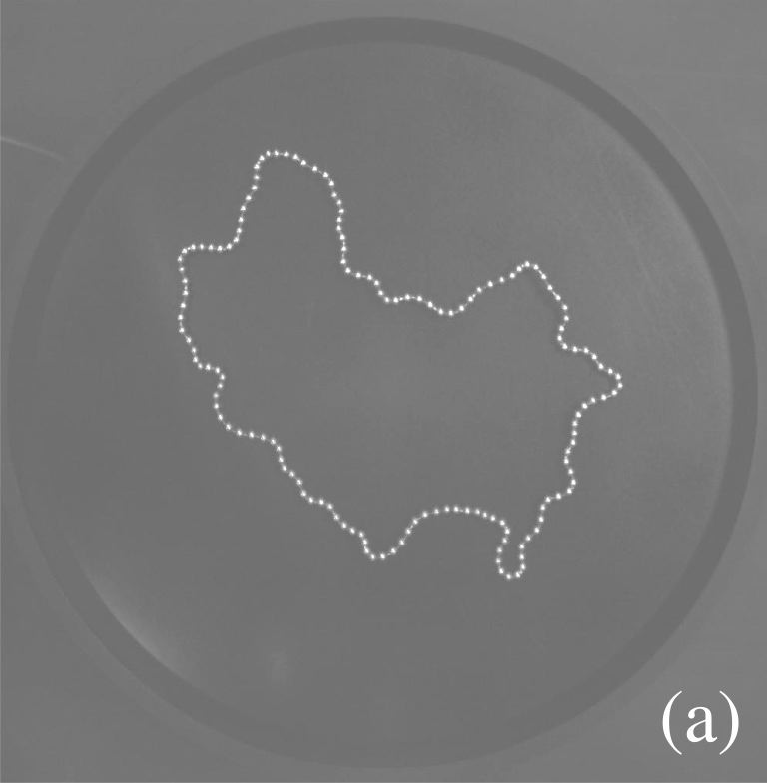}
\vspace{0.1cm}
\includegraphics[height=7.3cm]{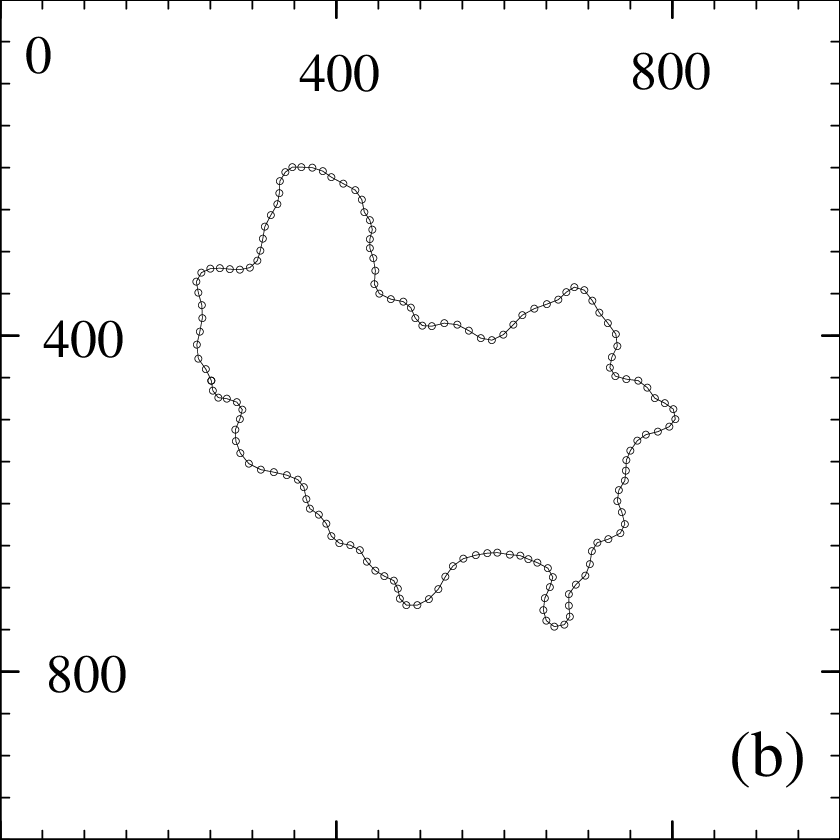}
\vspace{0.1cm}
\includegraphics[height=6.2cm]{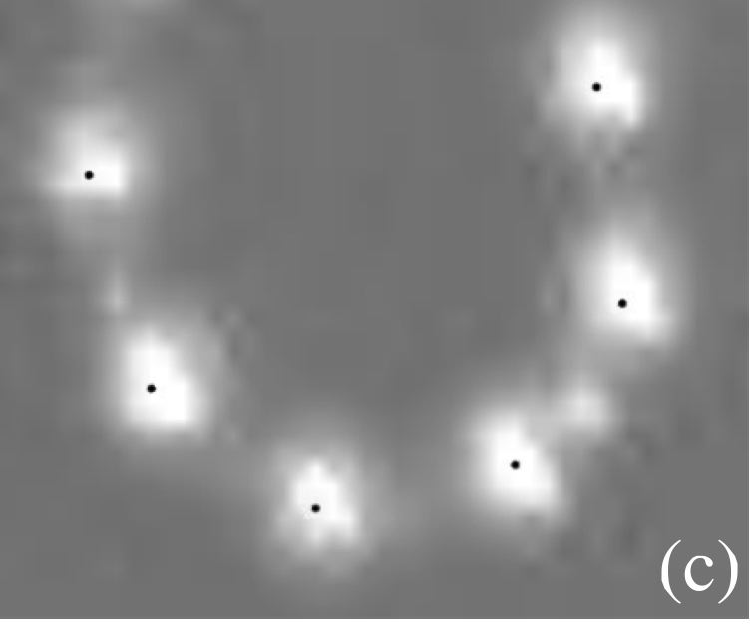}
\caption{\label{image_fig}
(a) Raw data image, (b) reconstruction by identifying the bead coordinates
(in pixels) and order, and (c) overlay of a section of the raw image and the
position data $(\bullet)$ for the center of the beads. Locally bright regions
between beads in (c) correspond to reflection from an interconnecting rod.}
\vspace{-0.0cm}
\end{figure}

Our data consist of digital images of the granular ring on the plate. We
illuminate the chain using approximately normally-incident light from far above
the plate. To minimize reflection from the background, we paint the acryllic
boundary black. Additionally, before black-anodizing the flat surface of the
plate is roughened by sand blasting. The very shallow surface inhomogeniety
helps randomize even simple collisions with the plate and diffusively scatters
far-field light. The poorly reflective background reduces noise in the digital
image. The normal far-field illumination gives an almost uniform reflected light
intensity across the plate area. Furthermore, the light reflected by the beads
is primarily from the center of the beads. Non-normal lighting preferentially
illuminates crescent-shaped areas on the surface of the beads. A representative
raw image is shown in Fig.~\ref{image_fig}a.

Typically, our images are acquired at sampling rates of $0.25-16$ Hz and have
a resolution of approximately $1000 \times 1000$ pixels. Depending on the
data set we have between $8.8$ and $13.0$~pixels to the bead diameter $d$.
The reflected light is brightest from the the center of the beads and
diminishes radially outwards. A sequence of standard image analysis
procedures such as background subtraction, intensity rescaling,
high-pass thresholding and smoothing result in reducing the data into roughly
$N \times 7 \times 7$ non-zero pixels for $N$ beads. Local maxima in
the intensity field are then determined and ranked by their peak intensity.
The intensity field of each set of $5 \times 5$ pixels centered about a
local maximum is fit to an azimuthally symmetric Gaussian. The fit parameters
of the position of the Gaussian peak are taken as the sub-pixel coordinates of
the bead. The fit to the standard deviation of the Gaussian gives an estimate to
the positional accuracy of about $0.05$ bead diameters. Figure~\ref{image_fig}c
shows a blow up of part of the raw image in Fig.~\ref{image_fig}a and
our determination of the bead centers.

In most of the images, our data extraction procedure resulted in more than $N$
local maxima. Most of these occur when a rod appears brighter than a bead. A
rod between two beads is visible in Fig.~\ref{image_fig}c. The frequency of this
occurrence is about $1\%$ of the number of beads. Since the distance between the
position of a rod and an adjacent bead is smaller than the minimum distance
between two adjacent balls, we eliminated these by iteratively applying a strict
minimum distance criterion.

The next step is to order the $N$ coordinates so that the ring can
be reconstructed. The ordering algorithm uses the geometrical
restrictions imposed by the interconnecting rods. Given two adjacent beads, the
third is searched for in the sector defined by the maximum inter-bead
separation and angle. This efficient greedy algorithm requires $N^2$ operations
to order an $N$ bead chain. The overall accuracy of our data extraction process
exceeds $96\%$ ,{\em i.e.}, in less than $4\%$ of the images are we unable
to reconstruct the chain. There are various reasons for this. The beads
are constructed by crimping sheet metal and so have a seam running along one
half of a great circle. The seam is comparatively duller in reflected light than
any other part of the beads and so in a few images our extraction process misses
these beads. Since the images are two-dimensional projections of the chain, the
strict inter-bead distance criteria can appear to be violated. In these
cases our algorithms may allow a false bead position. In Fig.~\ref{image_fig}b,
we show a complete reconstruction of a chain from the raw image data in
Fig.~\ref{image_fig}a.  Given the small number of misidentifications, we are
confident that these issues have a small effect on the results reported below.

\section{Local Properties}
\label{local_properties}

In this Section we present our results on the local properties of vertically
vibrated granular rings. Local properties are relevant to the scale of the bead
and here we discuss the inter-bead separation, the inter-rod angle, the average
curvature and the bead dynamics. We noted earlier that the chains are
constructed out of hollow shells that are
connected by rods. This connection imposes constraints on the minimum and
maximum separation between beads, on the maximum bending that 3 adjacent
beads can sustain, and on the displacements that occur on excitation.

\subsection{Inter-bead separations}

The inter-bead separation $\delta = |{\bf r}_{i+1} - {\bf r}_i|$ is measured
between every two consecutive bead positions for brass and nickel rings.
${\bf r_i}$ is position of bead $i$ in two dimensions. Hence,
there are $N$ measurements of $\delta$ for an $N$-bead ring. Our data is based
on no fewer than $5200$ measurements of $\delta$ for $N=13$ and more than $10^5$
samples for $N \geq 256$. From these data, we construct the probability
distribution function of the inter-bead separations $P(\delta)$. 

\begin{figure}
\includegraphics[height=6.5cm]{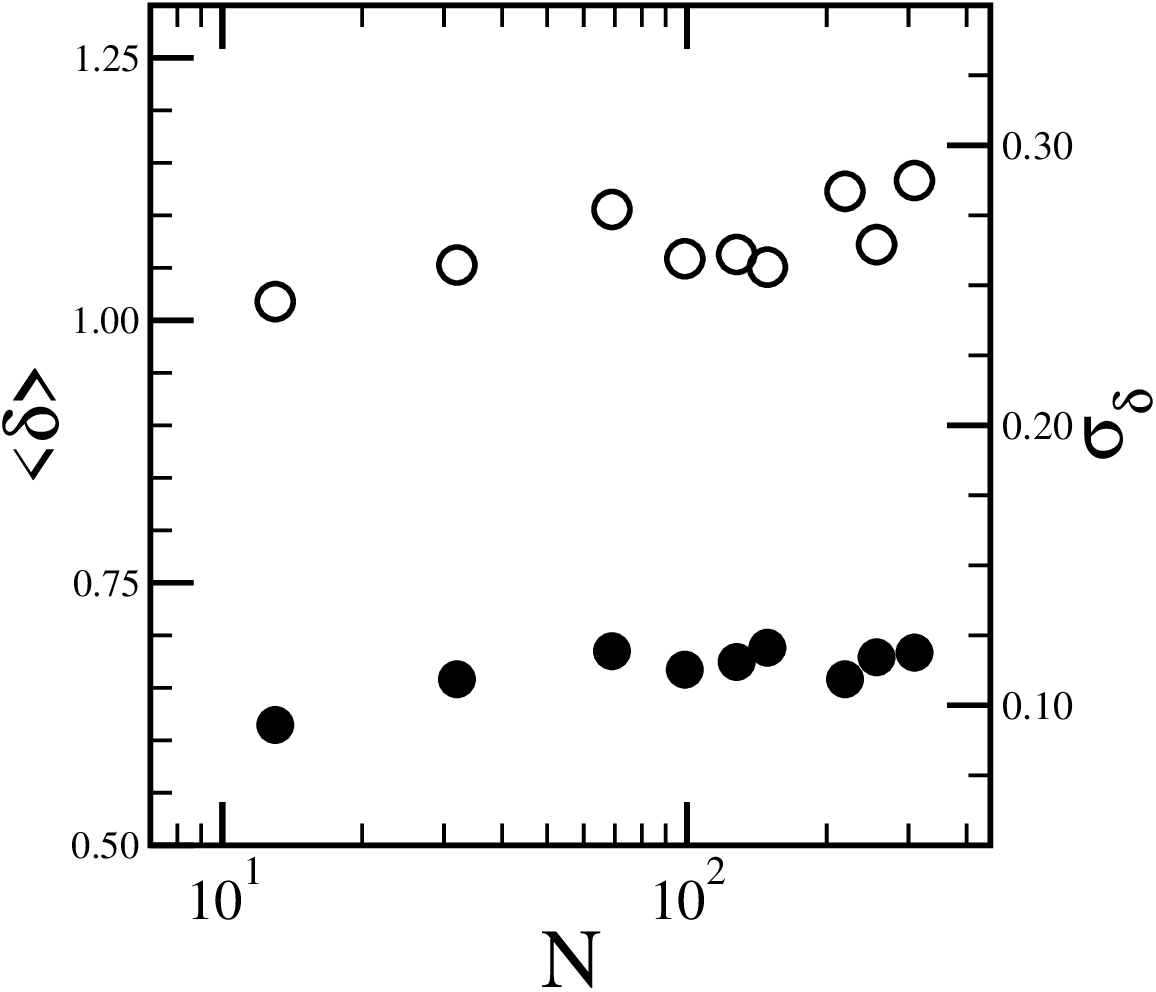}
\caption{\label{distances_fig}
The mean $\langle \delta \rangle (\circ)$ and standard deviation $\sigma_\delta
(\bullet)$ of the inter-bead separations $\delta$ vs ring length  $N$.}
\end{figure}

\begin{figure}
\includegraphics[height=6.5cm]{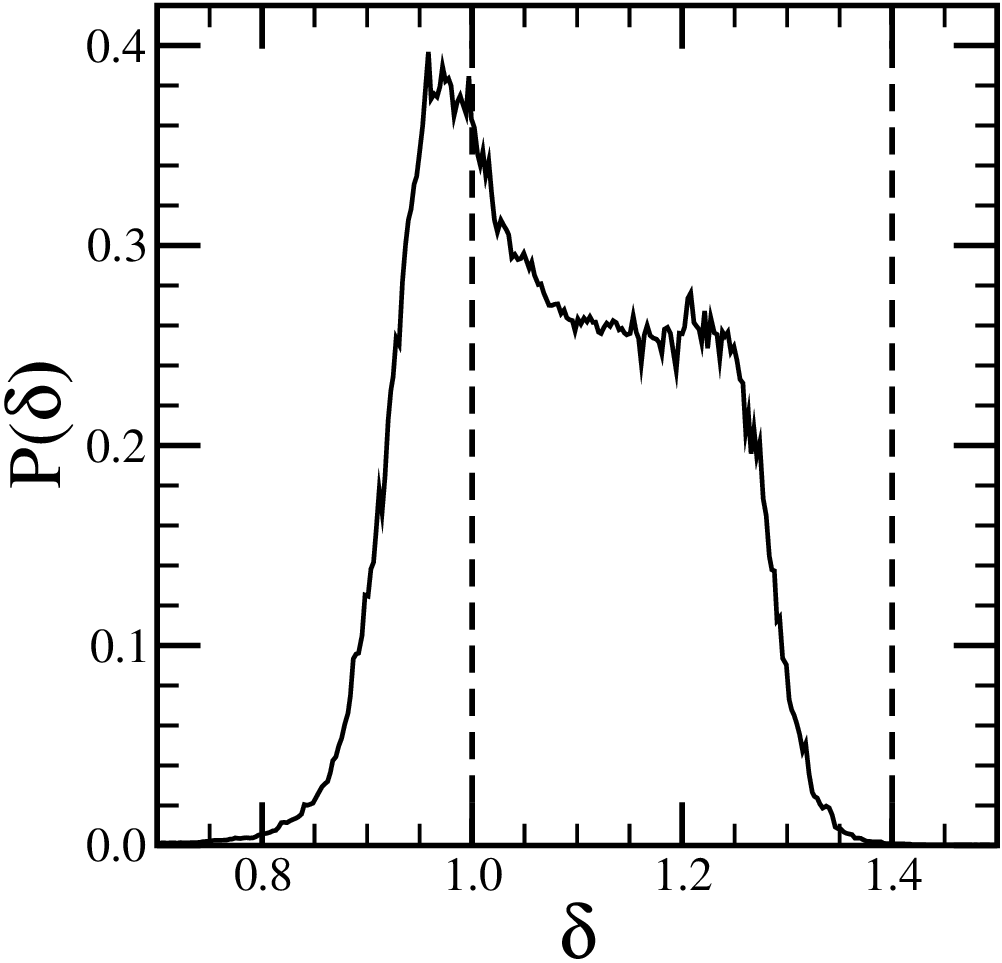}
\caption{\label{distances_pdf_fig}
The probability distribution $P(\delta)$ for the inter-bead separations. The
distribution combines data for rings with lengths $30 \leq N \leq 309$. The
dashed line $\delta =1 (1.4)$ in the theoretical lower (upper)  bound.}
\end{figure}

In Fig.~\ref{distances_fig}, we plot the mean $\langle \delta \rangle$
and standard deviation $\sigma_\delta$ as a function of $N$. For $N$
sufficiently large, $\langle \delta \rangle$ and $\sigma_\delta$ are
independent of $N$. The distributions $P(\delta)$ are similar in gross
feature for the brass and nickel rings for the various lengths $N$. Combining
the data for $30 \leq N\leq 309$, we obtain the distribution function shown
in Fig.~\ref{distances_pdf_fig}. The distribution has a sharp cut-off at
$\delta \approx 1.4$ corresponding to the maximal extension between the beads. A
much softer small $\delta$ cut-off appears at $\delta \approx 0.8$. The
minimal distance in an accurately two-dimensional chain would be at $\delta =
1$. The quasi-two-dimensional motion of the chain and our measuring the motion
on a planar projection allows values of $\delta < 1$. The peak of the
distribution at small separations is possibly due to bead-bead frictional
effects. Between the small-$\delta$ peak and the upper-$\delta$ cut-off we find
that the distribution is roughly flat. These intermediate separations occur
with significant and approximately equal probability, perhaps because of
the constant rod-bead friction at these extensions.

\subsection{Bending angles}

\begin{figure}[b]
\includegraphics[height=6.5cm]{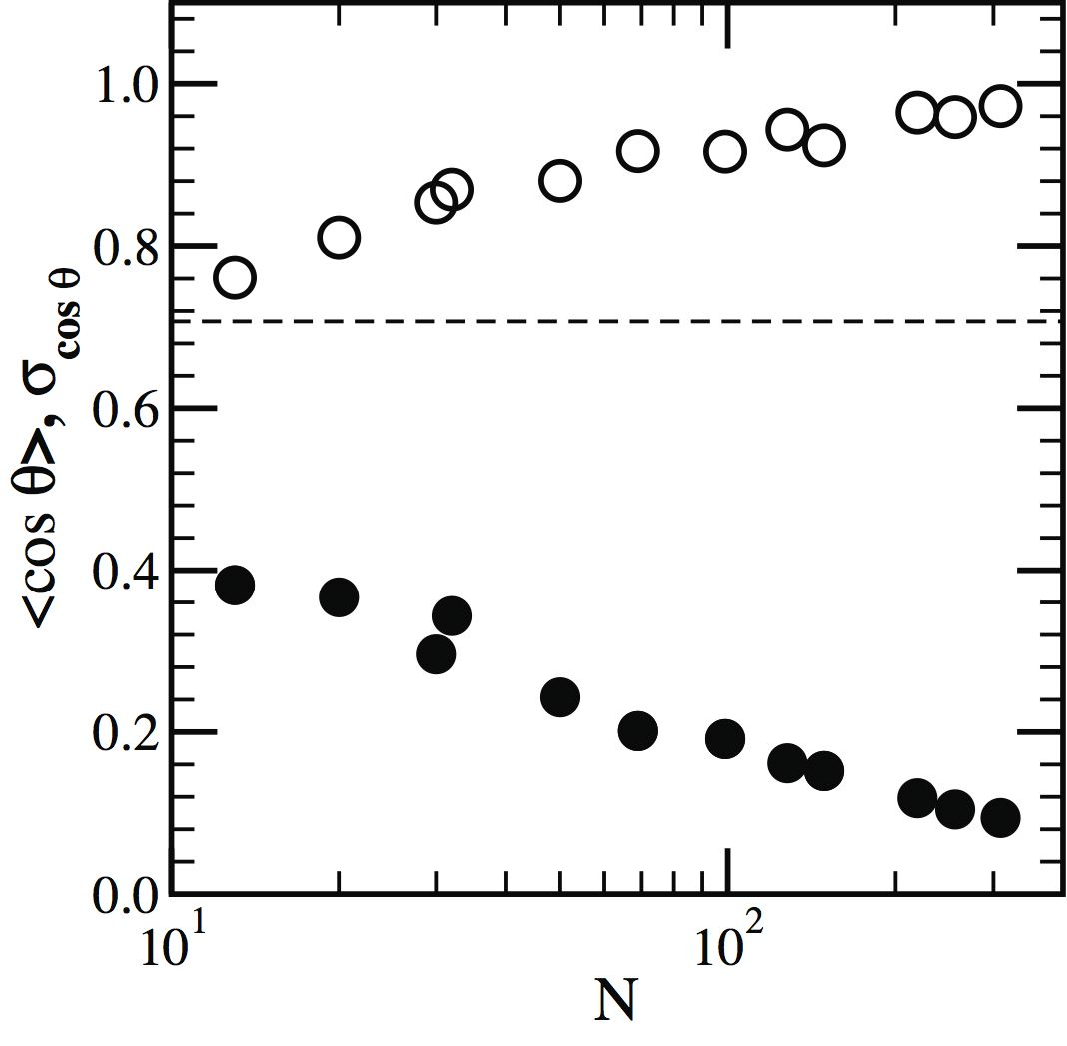}
\caption{\label{angles_fig}
The mean $\langle cos~\theta \rangle$ ($\circ$) and standard deviation
$\sigma_{cos \theta}$ ($\bullet$) of the bending angle $cos~\theta$ vs ring
length $N$. The dashed line corresponds to the lower bound set for $\langle
cos~\theta \rangle$ 
by a ring with $N=8$.}
\end{figure}

Given the positional ordering ${\bf r}_{i-1},{\bf r}_{i},{\bf r}_{i+1}$ of
three consecutive beads in the ring, we describe the bending by
\begin{equation}
cos~\theta = \frac{d{\bf r}_{i-1} \cdot d{\bf r}_i}{|d{\bf r}_{i-1}| |d{\bf
r}_{i}|}\,,
\end{equation}  
where $d{\bf r}_i = {\bf r}_{i+1} - {\bf r}_i$. There are $N$ values
of $cos~\theta$ for every $N$-bead ring. As with the inter-bead separations,
we have approximately $5200$ measurements of $cos~\theta$ for $N=13$ and more
than $10^5$ samples for $N \geq 256$ from which we determine the
distribution function. In Fig.~\ref{angles_fig}, we plot the mean
$\langle cos~\theta \rangle$ and standard deviation $\sigma_{cos \theta}$ of the
bending distribution functions for granular rings as a function of $N$. As $N$
increases, $\langle cos~\theta \rangle \rightarrow 1$ and $\sigma_{cos \theta}
\rightarrow 0$. This large $N$ behavior implies that a long ring is primarily
composed of locally straight segments. For $N$ approaching the smallest ring
size of $8$ beads, $\langle cos~\theta \rangle \rightarrow 0.7$ consistent with
the tightest bending.

\begin{figure}
\includegraphics[height=6.5cm]{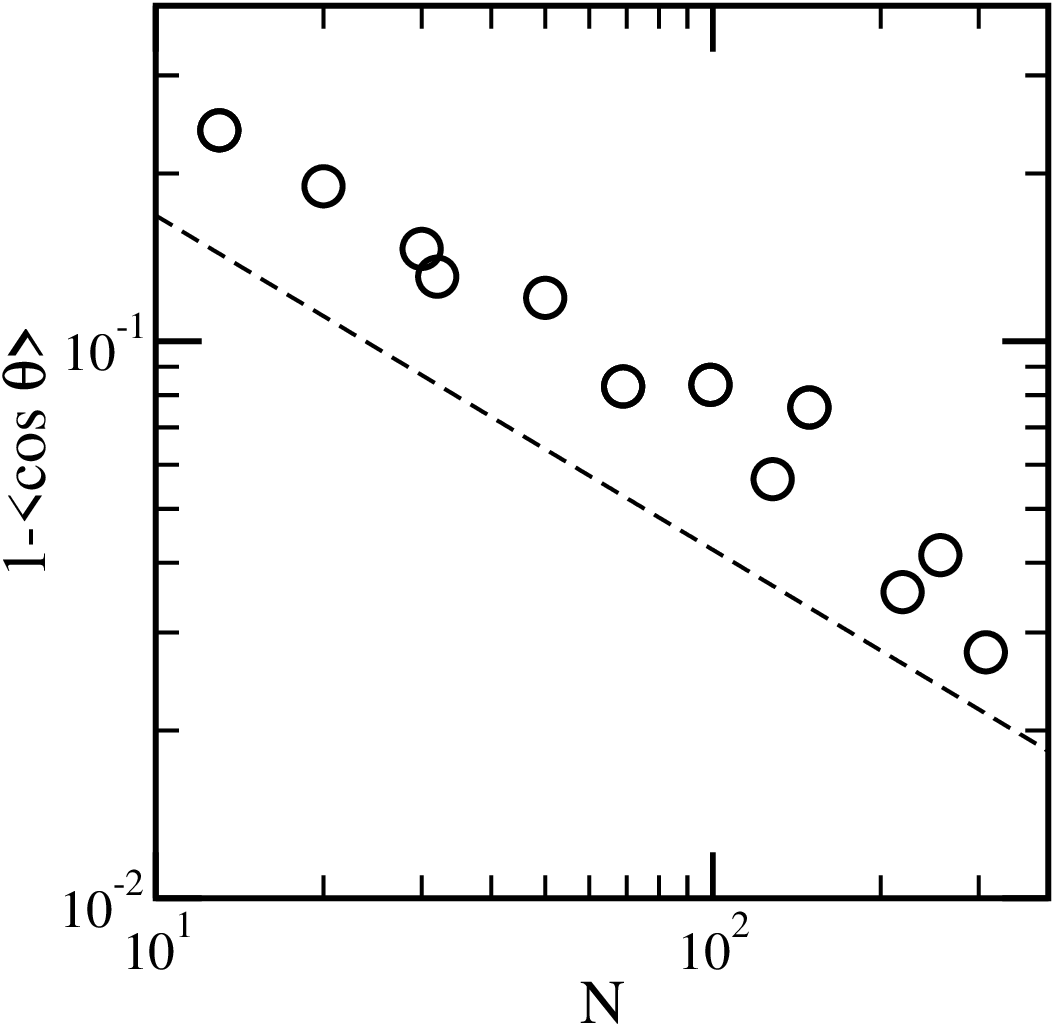}
\caption{\label{lim_angles_fig}
Plot of $1-\langle cos~\theta \rangle$ vs $N$. The dashed line is, up
to a constant, $N^{-0.6} $}
\end{figure}

The large $N$ limiting behavior of the mean bending angle is particularly
interesting. In Fig.~\ref{lim_angles_fig}, we plot the difference between the
limiting value of unity and the measured $\langle cos~\theta \rangle$. For an
ideal circular ring with $N$ beads, $1-\langle cos~\theta \rangle \sim N^{-2}.$
We find that although the granular ring becomes locally straight as its length
increases, it does so very differently from a circular ring. Our data show a
gentler approach with 
\begin{equation}
1-\langle cos~\theta \rangle \sim N^{-0.6}.
\label{eq:cosn}
\end{equation}  
This implies
that the chain curvature decreases slowly as the chain length increases.

\begin{figure}[b]
\includegraphics[height=6.5cm]{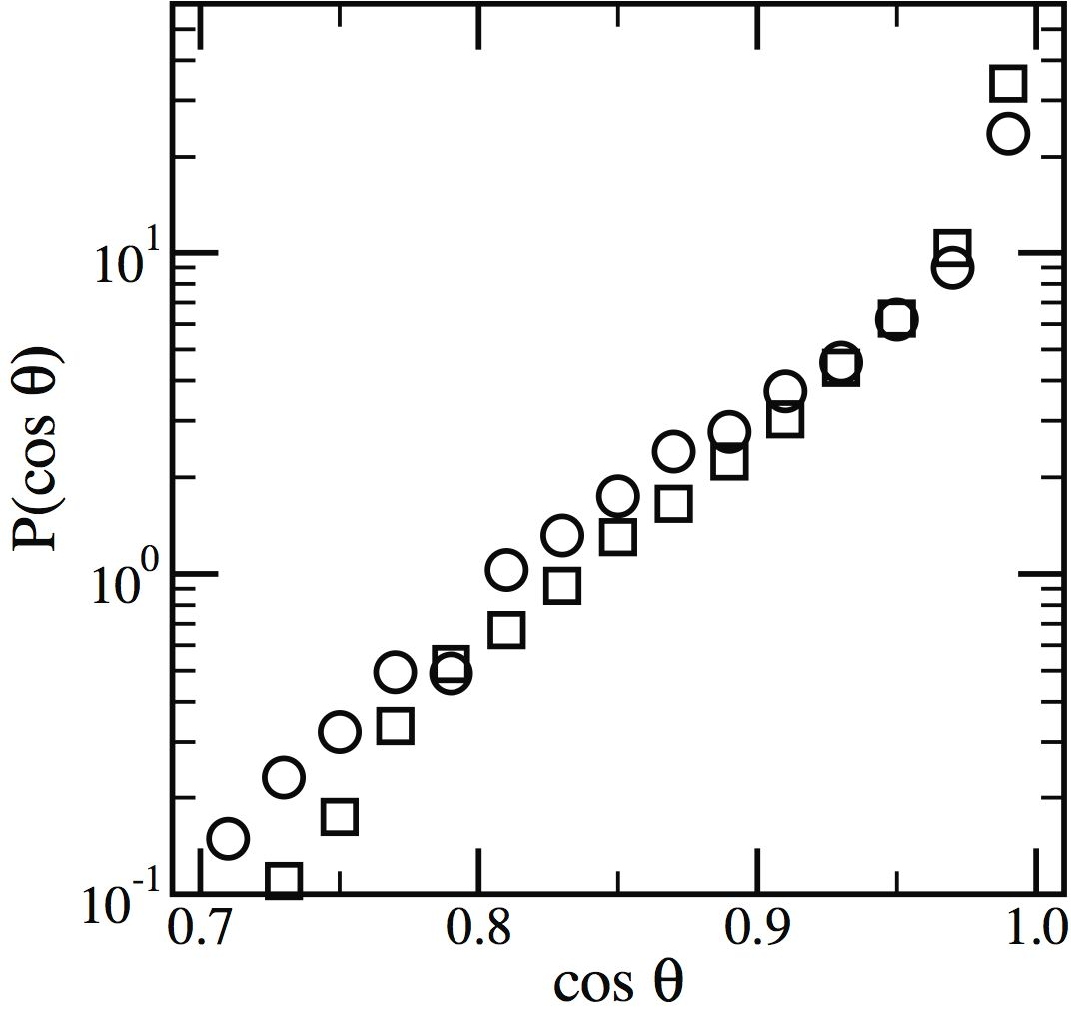}
\caption{\label{angles_pdf_fig}
The probability distribution of the bending angle $P(cos~\theta)$ for rings with
lengths $N = 32 (\circ)$ and $N=128 (\Box)$.}
\end{figure}

The bending distributions for representative small and large rings are shown in
Fig.~\ref{angles_pdf_fig}. The distributions peak near $cos~\theta \sim 1$ and
monotonically decrease for smaller $cos~\theta$. The peak and range depends
weakly on the number of beads $N$. The peak
probability for bending angles is higher for the longer rings whereas the range
of bending angles is larger for the shorter chains. There are significant
quantitative differences (not shown) between the distributions for brass and
nickel rings but the qualitative features are similar. The principal
difference is at large angles where the greater flexibility of the brass chains
increases the range of variation in $cos~\theta$.

\subsection{Average curvature}
\label{curvature}

The vibrated rings sample many microscopic configurations subject only to the
stretching and bending constraints imposed by the interconnecting rods and by
mutual self-exclusion. One of the properties of the configuration is the overall
extent to which the ring is curved. To probe this feature we use the ordered
positions to
compute the average squared curvature $K$ which we define as 

\begin{equation}
K = \frac{\oint {\bf k}(s)\cdot {\bf k}(s)~ds}{\oint {\bf k_{circle}}(s)\cdot
{\bf k_{circle}}(s)~ds}\,,
\label{K_eqn}
\end{equation} 
where ${\bf k}$ is given by
\begin{equation}
{\bf k}(s) = \frac{d^2 {\bf r}}{ds^2}; \hspace{5mm} {\bf r}(s) = (x(s),y(s))\,.
\label{k_def}
\end{equation}
In Eq.~\ref{k_def}, $s$ is the arc length along the chain and ${\bf r}(s)$ is a
unit-speed reparametrization of the chain. Then, by definition $|d{\bf
r}/ds| \equiv 1$ so ${\bf k}(s)$ measures the way the chain turns. ${\bf
k_{circle}}(s)$ is obtained by computing Eq.~\ref{k_def} for a circle with
circumference equal to the arc length of the chain. The
local squared curvature is $k(s)={\bf k}(s)\cdot {\bf k}(s)$. Averaging
over the ring we get $\oint k(s)~ds/\oint ds$. Similarly for the circle with
circumference equal to the chain's length, we have an average squared
curvature $\oint k_{circle}(s)~ds/\oint ds$. Normalizing the average squared
curvature of the ring by that of an equal arc length circle, we get the
quantity $K$ defined in Eq.~\ref{K_eqn}.

\begin{figure}
\includegraphics[height=6.5cm]{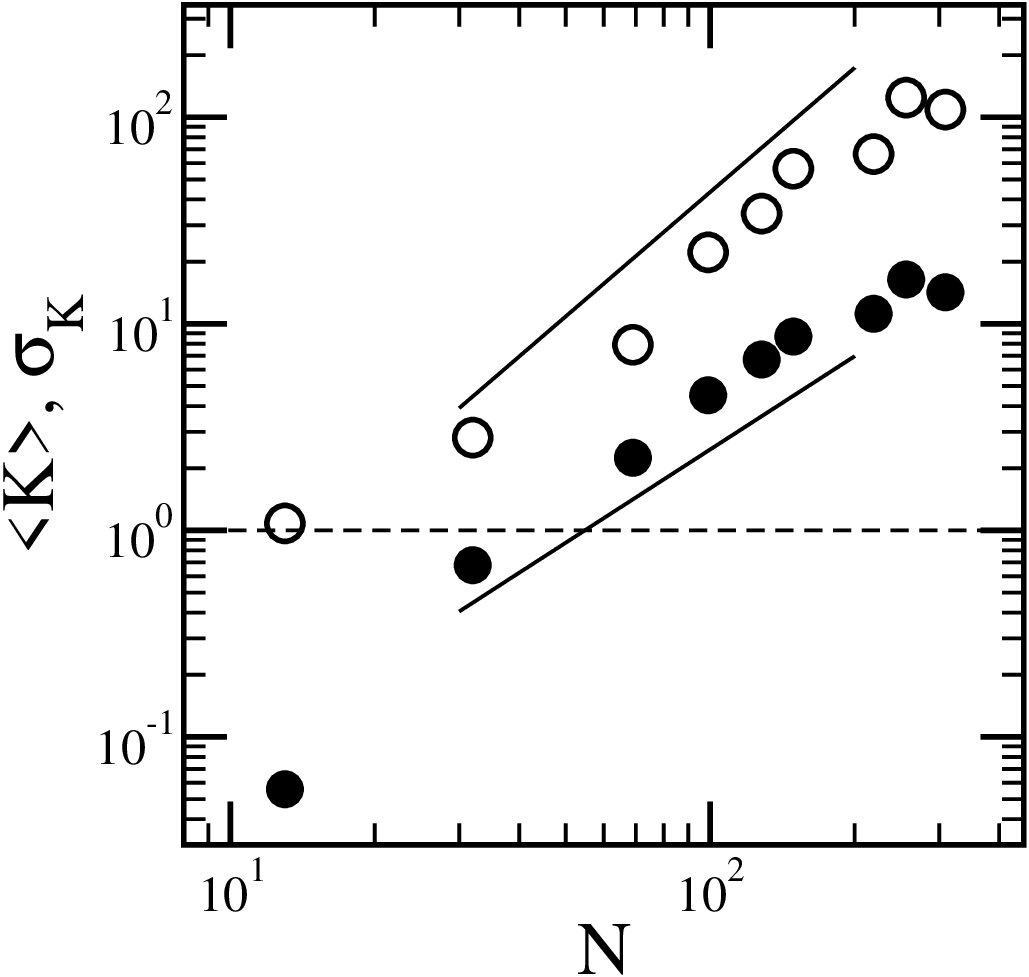}
\caption{\label{curvature_fig}
The mean $\langle K \rangle$ $(\circ)$ and standard deviation $\sigma_K$
($\bullet$) of the squared curvature, $K$ vs ring length $N$. The horizontal
dashed line $K=1$ represents a circle. The solid lines are proportional to
$N^{1.5}$ and $N^2$.}
\end{figure}

As a procedure, for each ring, we first calculate its circumference by adding
the separations between adjacent beads. We then reparameterize the positions of
the beads by the arc length $s$ so that ${\bf r}_{i}=(x_i,y_i) =
(x(s_i),y(s_i))$ such that the ring has unit circumference and is traversed with
unit speed. Since $\oint {\bf k}_{circle}(s)\cdot {\bf k}_{circle}(s)~ds =
4\pi^2$ for a circle of unit length and speed, we have normalized $K$ such that
as the ring approaches a circle $K \rightarrow 1$. Hence $K$ measures the
departure from a circular shape. In practice, the integral in Eq.~\ref{K_eqn} is
replaced by a summation over the $N$ discrete bead positions and ${\bf
k}(s_{i})$ is approximated by the difference
\begin{equation}
{\bf k}(s_{i}) = \frac{d^2 {\bf r}_{i}}{ds_{i}^2} = \frac{d{\bf r}_{i+1}-d{\bf
r}_{i}}{\frac{1}{2}(s_{i+1}-s_i)+\frac{1}{2}(s_i-s_{i-1})}.
\label{ki_def}
\end{equation}

Using in excess of $6800$ images or approximately $425$ realizations per chain,
we have measured the mean and rms values of $K$ for brass and nickel rings for
several $N$. In Fig.~\ref{curvature_fig}, we plot the mean $\langle K
\rangle$ and standard deviation $\sigma_K$ as a function of $N$. The two
smallest rings have $N=13$ and $20$. They are approximately circular and tightly
constrained, resulting in $\langle K \rangle \approx 1$;  the smallest ring has
$N=8$ and is practically rigid. For larger $N$ we find that the scaling exponent
$\mu\sim 2$ in the power-law $\langle K \rangle \sim N^\mu$ for $20 \leq N
\leq 256$. This implies, in agreement with Eq. 2,  that as $N$ increases, the
ring increasingly deviates
from a circle. The standard deviation $\sigma_K$ from the data shows a scaling
relation $\sigma_K \sim N^\lambda$ with $\lambda \sim 1.5$. This observation
is consistent with Eq.\ \ref{eq:cosn}.

\subsection{Bead displacements}
\label{bead_displacements}
The dynamics of the beads are studied by measuring the displacement of a bead
as a function of time. It is well known that the mean square displacement of a
randomly excited particle scales linearly with time while ballistic motion
results in quadratic scaling. It is interesting to probe how a bead
constrained to the ring behaves.

We restrict our analysis to images acquired at $16$~Hz which is sufficiently
fast so that bead positions between two consecutive images show small changes.
Consequently, we can identify beads in an image with the preceding image.
Discerning the bead identity is increasingly more difficult with slower image
acquisition rates since the bead displacements are quite large. The ordered
position data of the beads in the ring is obtained as described earlier
in Section~\ref{data_extraction}. Then assuming the bead identity in the first
image, we order the bead positions of the second image so that they correspond
to the preceding image. Since there is only $1$ plate oscillation and therefore
on average $1$ chain excitation between consecutive images, we
are confident that we have accurately identified the beads. Additionally, we
visually verify the ordering for a subset of the data sets. We implement our
bead identity algorithm by translating the data from both images to a
common center of mass and then minimizing the mean square position difference
between the two images up to cyclic permutations and orientation of the second
image. The procedure is iterated until all the images are properly ordered.

\begin{figure}
\includegraphics[height=6.5cm]{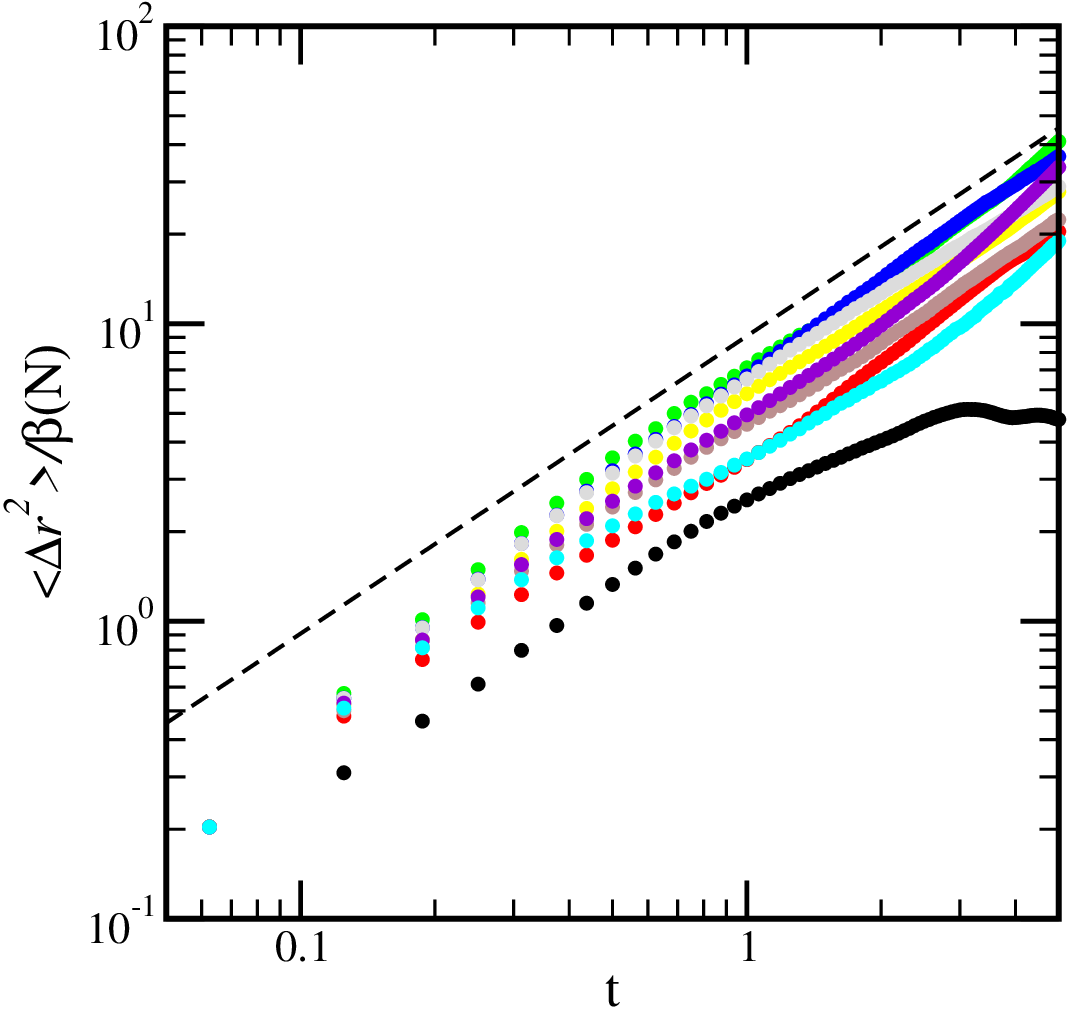}
\caption{\label{displacements_fig}
The mean squared displacements $\langle \Delta r^2 \rangle$ of the beads as a
function of time $t$ for various ring length $N$. The displacements are scaled
by a ring length-dependent constant $\beta(N)$ so that all $\langle \Delta r^2
\rangle$ are the same for the shortest experimentally observed time $t=1/16$ s.
The dashed line is proportional to $t$. The $\langle \Delta r^2 \rangle$ for the
smallest ring $N=13$ is shown in black solid circles.}
\end{figure}

The squared displacement that bead $i$ in a ring with $N$ beads undergoes
between time $\tau$ and $\tau + t$ is $\Delta r_{i}^2(t,\tau,N) = ({\bf r_i(\tau
+ t)}-{\bf r_i(\tau)})\cdot({\bf r_i(\tau + t)}-{\bf
r_i(\tau)})$. For the data, the times $\tau$ and $t$ are chosen to
coincide with the image acquisitions, {\em i.e.}, multiples of $1/16$ s, and the
displacement is measured in units of the bead diameter. Note that both images
are translated to the origin, {\em i.e.}, the center of mass is
subtracted from the bead coordinates before calculating the squared
displacement.

Averaging over all beads $i$ in the ring and over possible values of $\tau$, we define
the mean squared bead displacement:
\begin{equation}
\langle \Delta r^2 \rangle (t,N) = \langle \Delta r_i^2 (t,\tau,N)
\rangle_{\tau,i}\,.
\end{equation}
We probe the time dependence of $\langle \Delta r^2 \rangle$ by assuming a
length dependent constant factor $\beta(N)$ such that the mean squared
displacements are the same for the shortest experimentally observed time $t=1/16$
s. In Fig.~\ref{displacements_fig}, we plot $\langle \Delta r^2
\rangle/\beta(N)$ versus $t$. For times corresponding to about 10 plate
oscillations, $\langle \Delta r^2 \rangle/\beta(N)$ increases linearly with
time for all rings. This behaviour is consistent with diffusive motion. Thus,
although the
ring is mechanically and athermally driven, collective properties may still
behave as if
they are randomly (thermally) forced \cite{ebn_01,hastings_02}.  
For shorter times the mean squared displacements increase faster than
linearly with time for all but the smallest ring which is consistent with the
expected crossover from very short time ballistic motion to long time diffusive
dynamics.

\begin{figure}
\includegraphics[height=6.5cm]{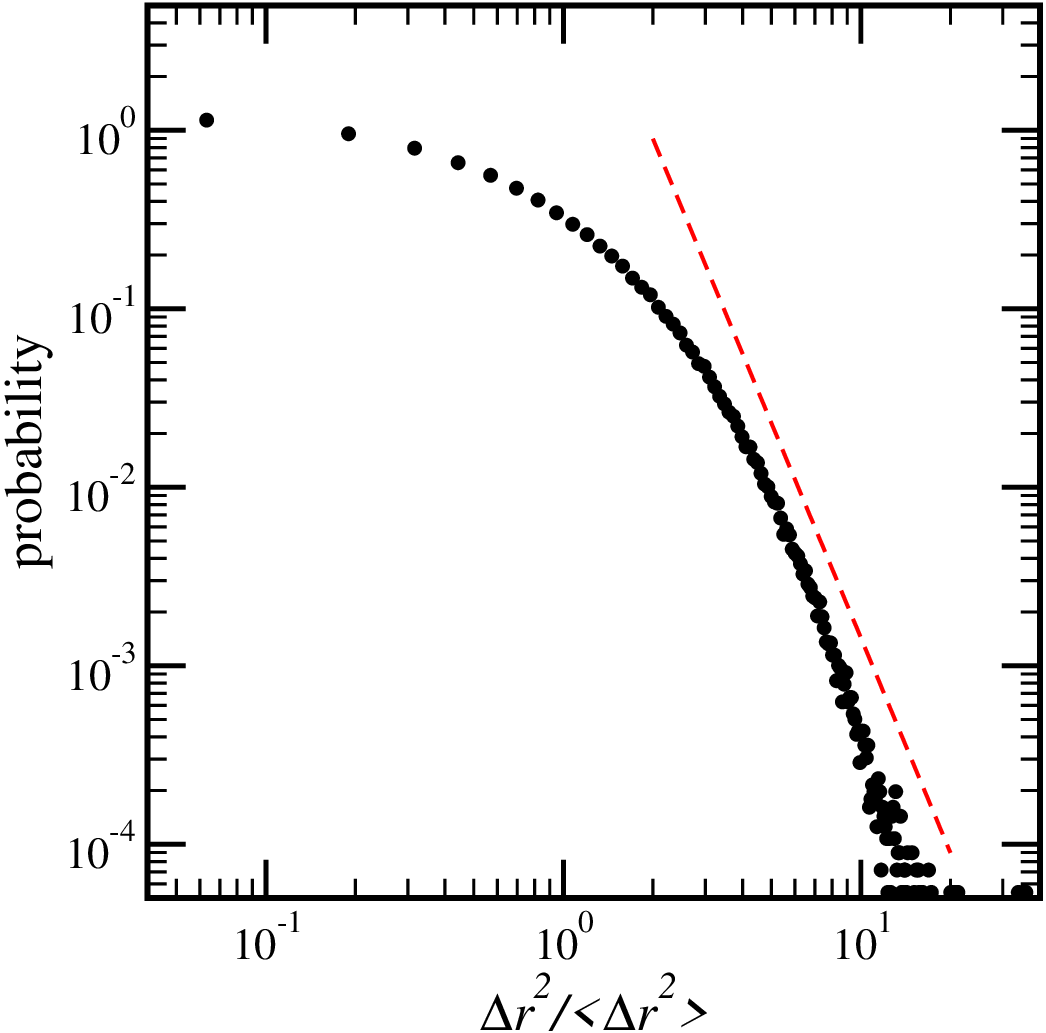}
\caption{\label{displacements_pdf_fig}
The probability distribution function for the $\Delta r^2$/$\langle \Delta r^2
\rangle$ at $t=1/16$ s for all ring lengths $N$. The dashed line has slope $-4$.
}
\end{figure}

For the shortest observation interval of $t=1/16$ s, we study the distribution
of the squared displacements. Since each bead in the
ring collides with the plate, on average, once per oscillation, it is highly
probable that several beads in the chain will have altered
direction due to a collision. Thus, it is unlikely that the $1/16$ s interval is
sufficiently short to deduce an instantaneous velocity. Since the
bead-plate collisions occur over a range of phase in the oscillation, however,
the beads
encounter a range of displacements. Large displacements correspond to collisions
and smaller displacements to beads being dragged along by the chain
connectedness. In Fig.~\ref{displacements_pdf_fig}, we plot the
probability distribution function of $\Delta r^2$ measured at $t=1/16$ s. The
squared displacement is normalized by its mean value. Our data show that small
squared displacements are highly probable decaying slowly until $O(\langle
\Delta r^2 \rangle)$. For larger displacements the probability decays more
rapidly and the large displacement tail decay is consistent with $(\Delta
r^2)^{-4}$.

\section{Global Properties}
\label{global_properties}

Global characteristics of granular rings probe the overall conformation and
dynamics of the chain. Below we describe the conformation, roughly speaking, by
the chain's size. The size, defined to be the mean radius of gyration,
is obtained by averaging over many observations. The global dynamics are
described by the motion of the center of mass of the ring.

\subsection{Radius of Gyration}
\label{gyration}

A measure of the global size of a long chain or a polymer is its radius of
gyration. Since the chain's configuration is constantly changing owing to
collisional excitation with the plate, better measures are the
statistical averages of radius of gyration. Given the $N$ position coordinates
${\bf r_i} = (x_i,y_i)$ for the beads $i  = 1,..,N$, we calculate the radius of
gyration $R_g$ from

\begin{equation}
{R_g}^2 = \frac{1}{N} \sum_{i=1}^{N} ({\bf r_i} - {\bf r_{cm}})^2\,,
\label{rg_eqn}
\end{equation}
where ${\bf r_{cm}} = (x_{cm},y_{cm})$ is given by 

\begin{equation}
{\bf r_{cm}} = \frac{1}{N} \sum_{i=1}^{N} {\bf r_i}\,.
\label{cm_eqn}
\end{equation}

Since our coordinates are in units of the bead diameter $d$, the radius of
gyration $R_g$ is in these units. We have measured $R_g$ for granular rings for
several values of $N$.  Since there are no self-intersections, the ring can be
likened to a self avoiding walk that returns to itself.

\begin{figure}
\includegraphics[height=6.5cm]{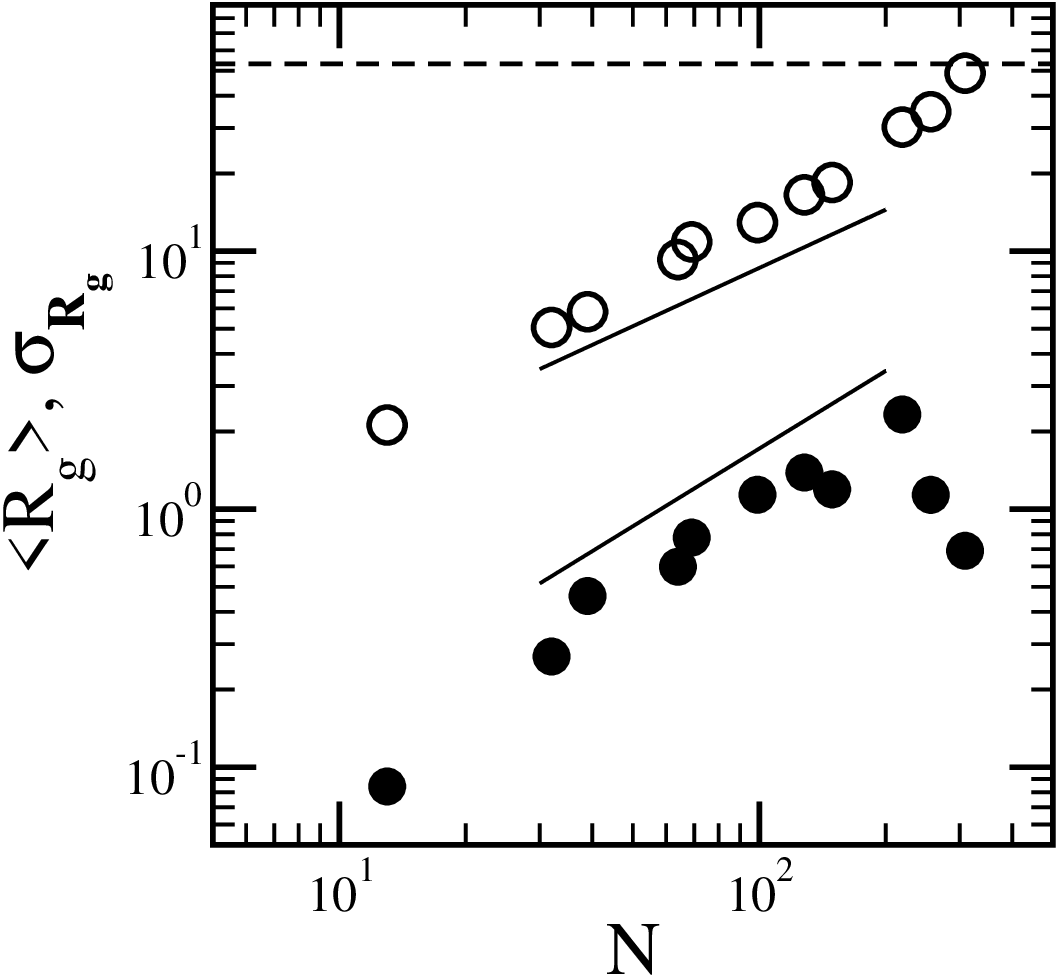}
\caption{\label{gyration_fig}
The mean $(\circ)$ and standard deviation $(\bullet)$ of the radius of gyration
vs the length of the rings $N$. The dashed horizontal line is the radius of the
plate. The two solid lines have slopes of $3/4$ (top) and $1$ (bottom),
respectively. }
\end{figure}

Averaging over approximately $450$ realizations per granular ring of length $N$,
we measure the mean and rms values of $R_g$. In Fig.~\ref{gyration_fig}, we
plot $\langle R_g \rangle$ and $\sigma_{R_g}$ as a function of $N$ for
$13 \leq N \leq 309$. In these experiments the lateral boundary
has a radius of $53.2$ bead diameters. This is an absolute upper bound on
$\langle R_g \rangle$ and is shown by the horizontal line in
Fig.~\ref{gyration_fig}. Our $\langle R_g \rangle$ data for nickel and brass
chains are consistent and increase with $N$. A power-law fit for the range $20
\leq N \leq 219$ gives an exponent $\nu = 0.85 \pm 0.05$. The $\sigma_{R_g}$
data in Fig.~\ref{gyration_fig} shows a linear increase with $N$ for $20 \leq N
\leq 219$ with a sharp drop for larger $N$. The sharp drop is associated with
the constraining effect of the lateral boundary. 

Self avoiding walks have universal scaling characteristics in the large-$N$
limit. In particular, $\langle R_g \rangle \sim N^\nu$ with Flory exponent
$\nu={3\over4}$ in two dimensions \cite{degennes,doi-edwards}.
Experiments with surface adsorbed DNA\cite{maier_99}, polymer molecular
monolayers\cite{vilanove_80} and connected plastic spheres\cite{prentis_02} have
measured exponents consistent with the 2D Flory value $\nu={3\over4}$. Both the
DNA and plastic chain had free ends and by being constrained to 2D were
self-avoiding. Our measured exponent $\nu = 0.85 \pm 0.05$ is for a ring
(likened to self avoiding walk that closes on itself) and is
somewhat higher though not inconsistent with these experiments. The
$\sigma_{R_g}$ increase linearly with $N$ in the range $20 \leq N \leq 219$. 

\subsection{Center of Mass Motion}
\label{cm_motion}

Under oscillation the rings sample many configurations with different positions
of their center of mass. Since the plate is accurately level, the rings do not
drift gravitationally. Any motion of the center of mass is then associated with
the overall mobility of the rings. We study the motion of the center of mass of
nickel rings with $13 \leq N \leq 309$. The data are acquired at $16$~Hz which
is sufficiently fast for probing the motion of the ring's center of mass. Given
the center of mass ${\bf r}_{cm}$ we calculate the magnitude of the average
velocity $v$ given by 

\begin{equation}
v = \sqrt{\frac{d{\bf r}_{cm}}{dt} \cdot \frac{d{\bf r}_{cm}}{dt} }\,.
\end{equation}

\begin{figure}
\includegraphics[height=6.5cm]{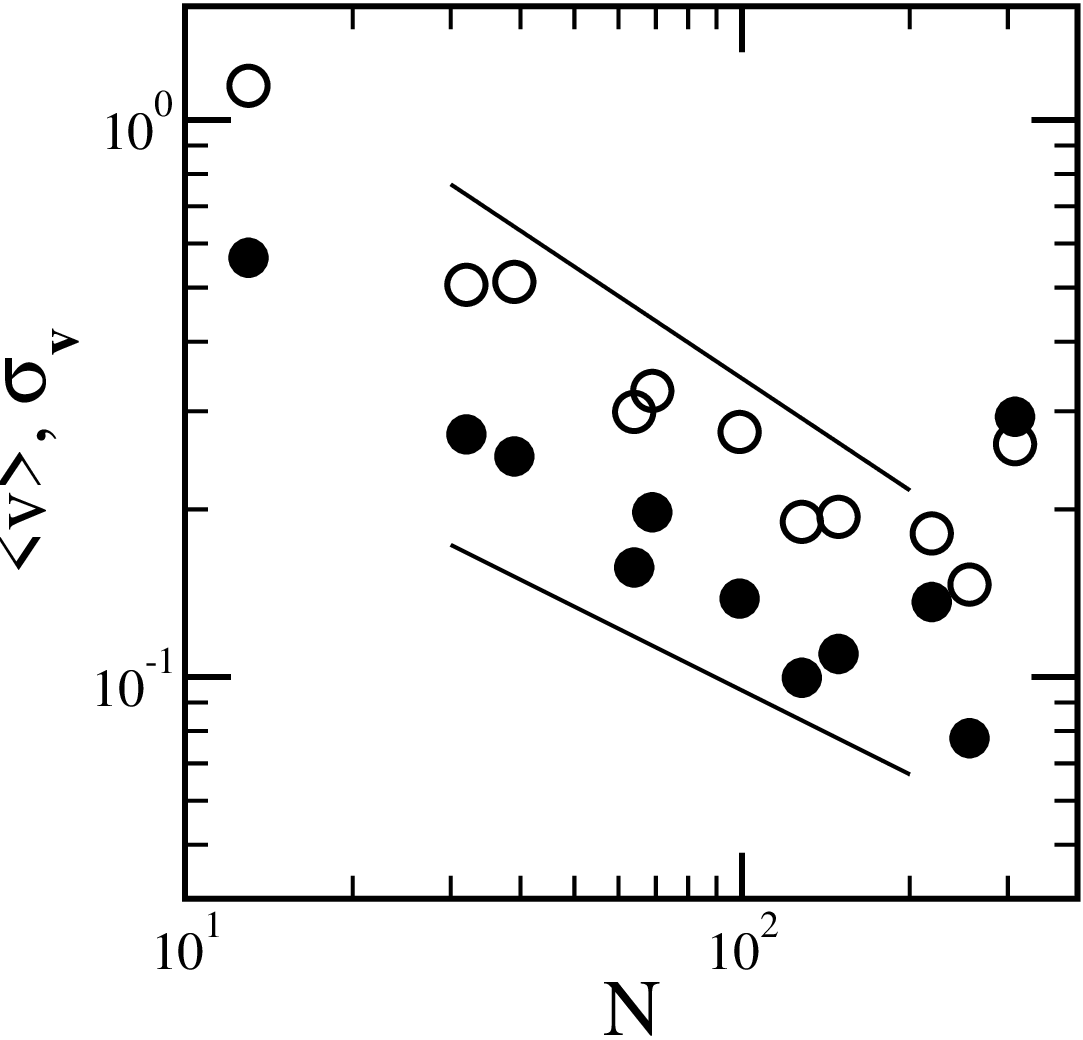}
\caption{\label{cm_fig}
The mean $(\circ)$ and standard deviation $(\bullet)$ of the center of mass
velocity. The straight lines are power-law fits to the data with slopes -0.67
(mean) and
-0.59 (standard deviation), respectively.}
\end{figure}

>From approximately $400$ images per ring we measure the mean and rms
values of $v$. In Fig.~\ref{cm_fig}, we plot the mean $\langle v \rangle$
and standard deviation
$\sigma_v$ for several $N$. For our largest ring $N=309$ significant interaction
with the lateral boundary appears to be responsible for the unexpectedly large
$\langle v \rangle$ and $\sigma_v$. Assuming scaling relations $\langle v
\rangle \sim N^\eta $ and $\sigma_v \sim N^\zeta $ power-law fits for $13 \leq N
\leq 256$ give $\eta = -0.67 \pm 0.08$ and $\zeta = -0.59 \pm 0.14$. The speed
$v$ varies from just over $1$ bead diameter/s for $N=13$ to about $1/10$
bead diameters/s for $N=256$. As $N$ increases more and more segments of the
ring move independently resulting in large cancelations to contributions to
center of mass motion.  Qualitatively, this diminished mobility with increasing
chain
length resembles the behavior observed for a random polymer where the
diffusivity
is inversely proportional to the molecular weight \cite{doi-edwards}.

\begin{figure}
\includegraphics[height=6.5cm]{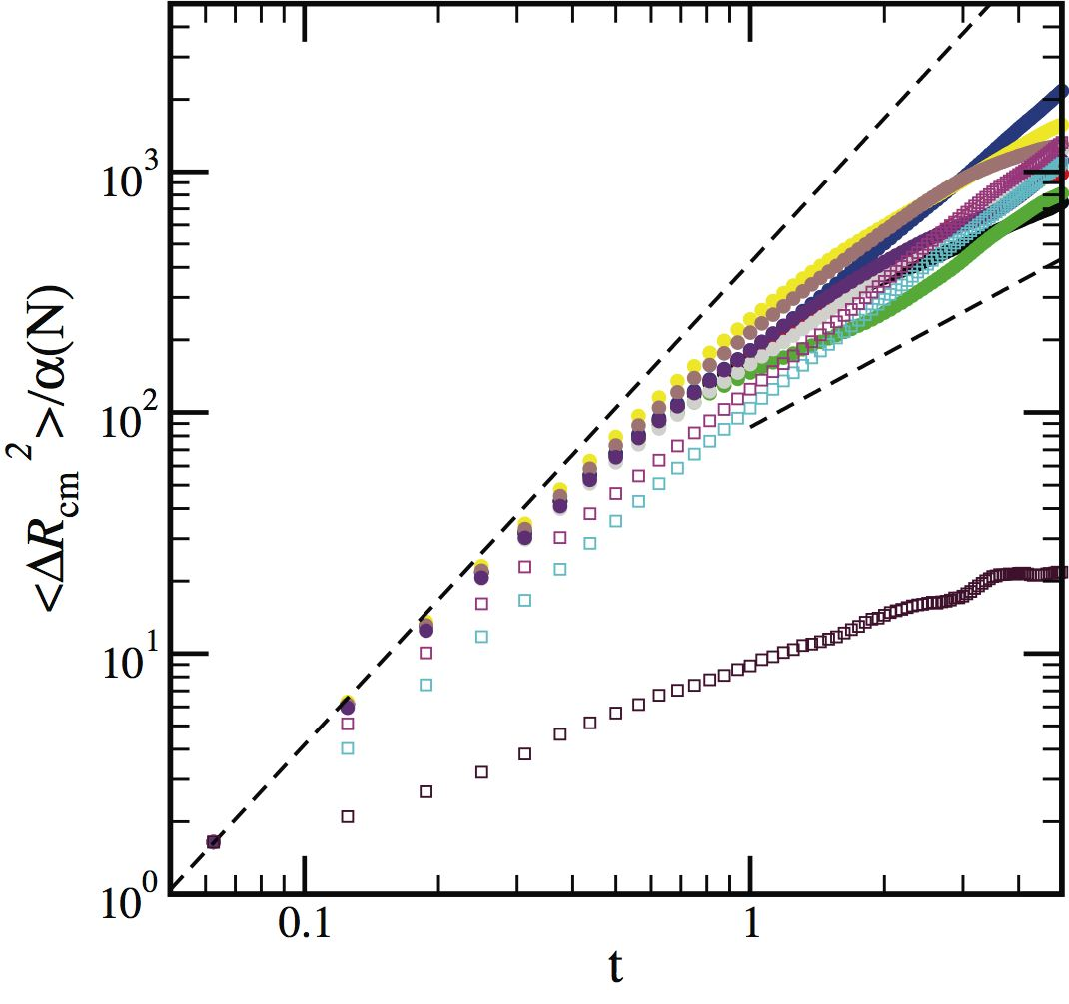}
\caption{\label{cm_displacements_fig}
The mean squared displacements $\langle \Delta R_{cm}^2 \rangle$ of the ring
center of mass as a function of time $t$ for various ring length $N$. The
displacements are scaled by an ring length-dependent constant $\alpha(N)$ so
that all $\langle \Delta R_{cm}^2 \rangle$ are the same for the shortest
experimentally observed time $t=1/16$ s. The dashed lines are proportional to
$t$ and $t^2$. The  $\langle \Delta R_{cm}^2 \rangle$ from the three longest
rings $N=219$, $256$ and $309$ are shown in magenta, cyan and maroon open square
symbols respectively.}
\end{figure}

As with the bead displacements studied in Section~\ref{bead_displacements}, we
can study the dynamics of the ring center of mass. The
squared displacement that ring center-of-mass undergoes
between time $\tau$ and $\tau + t$ is $\Delta R_{cm}^2(t,\tau,N) = ({\bf
r_{cm}(\tau + t)}-{\bf r_{cm}(\tau)})\cdot({\bf r_{cm}(\tau + t)}-{\bf
r_{cm}(\tau)})$. For the data, the times $\tau$ and $t$ are chosen to
coincide with the image acquisitions, {\em i.e.}, multiples of $1/16$ s and the
displacement is measured in units of the bead diameter. Averaging over all
possible values of $\tau$ we define the mean squared ring center
of mass displacement:
\begin{equation}
\langle \Delta R_{cm}^2 \rangle (t,N) = \langle \Delta R_{cm}^2 (t,\tau,N)
\rangle_{\tau}\,.
\end{equation}
The time dependence of $\langle \Delta R_{cm}^2 \rangle$ is plotted in
Fig.~\ref{cm_displacements_fig}. Here, we have assumed a length dependent
constant factor $\alpha(N)$ such that the mean squared displacements are the
same for the shortest experimentally observed time $t=1/16$ s.

For times corresponding to less than about 10 plate oscillations, the $\langle
\Delta R_{cm}^2 \rangle/\alpha(N)$ increase quadratically with time for all
rings except the three largest rings with $N=219$, $256$ and $309$. The
constraining effect of the plate's lateral boundary explains the less
than quadratic time dependence of $\langle \Delta R_{cm}^2 \rangle/\alpha(N)$
for large rings. For times in excess of about 10 plate oscillations, the $\langle
\Delta R_{cm}^2 \rangle/\alpha(N)$ approaches a linear increase with time. The
short time behaviour is consistent with ballistic motion of the ring center of
mass. This suggests that at short times the motion of the ring center of mass
is dominated by coherent large bead displacements. The longer time behaviour
averages over several, apparently random, large displacements resulting in the
approach to diffusive motion.

\begin{figure}
\includegraphics[height=6.5cm]{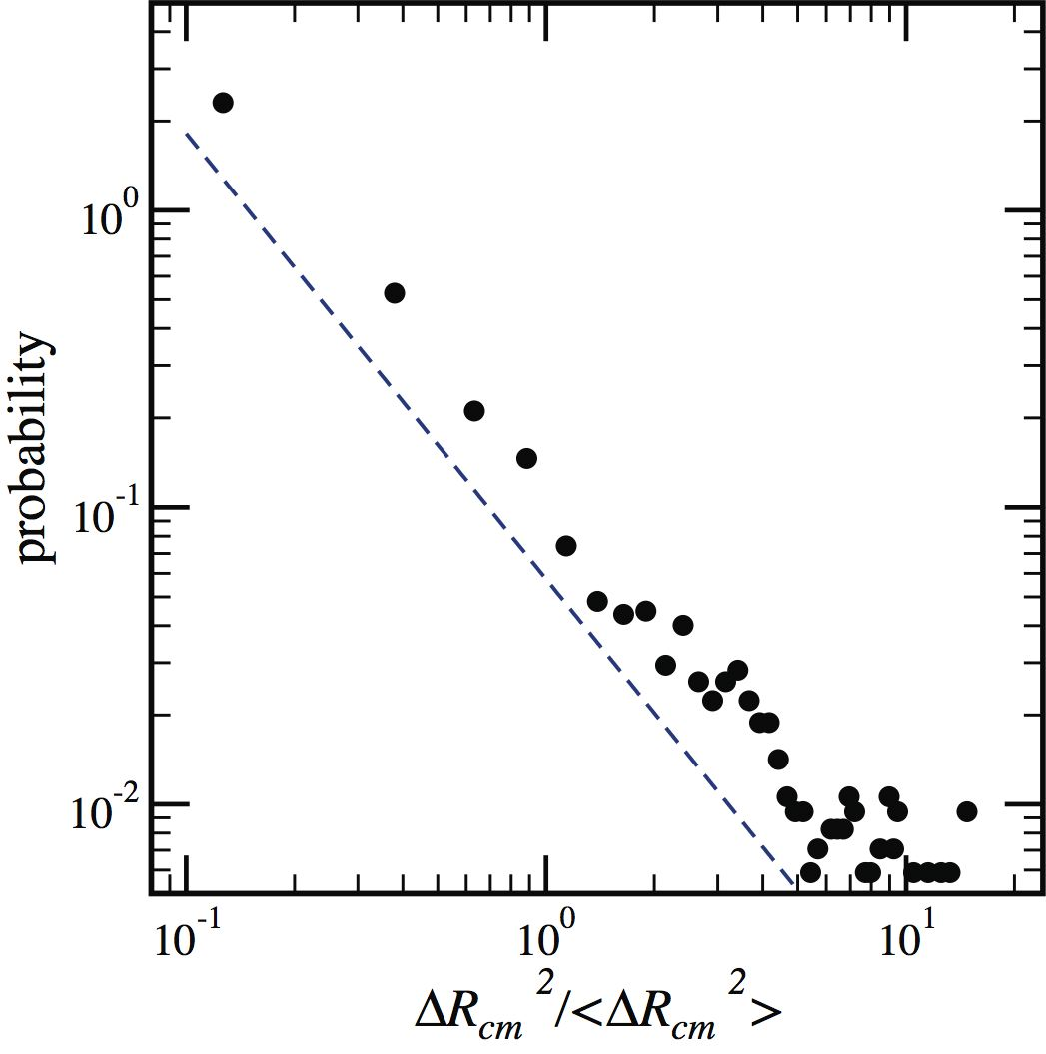}
\caption{\label{cm_displacements_pdf_fig}
The probability distribution function for the $\Delta R_{CM}^2$/$\langle \Delta
R_{CM}^2 \rangle$ at $t=1/16$ s for all ring lengths $N$. The dashed line has
slope $-1.5$.}
\end{figure}

The distribution of the squared displacements of the ring center of mass is
studied for the shortest observation interval of $t=1/16$ s. At this time
scale the center of mass motion is ballistic and so an average velocity
proportional to the displacement may be infered.
In Fig.~\ref{cm_displacements_pdf_fig}, we plot the
probability distribution function of $\Delta R_{cm}^2$ measured at $t=1/16$ s.
The squared displacement is normalized by its mean value. Our data show that
small squared displacements are highly probable with the probability
decaying consistent with $(\Delta R_{cm}^2)^{-1.5}$. The small squared
displacements of the ring center of mass arise from averaging the almost
independent motions of the beads on the ring. The less probable large
squared displacements occur from rarer coherent motions of segments of beads
on the rings.

\section{Conclusion}
\label{conclusion}

In this paper we presented an experimental study of vibrated granular
rings. Our study elucidates the statistical characterization of
various local and global properties. We have been
particularly interested in power-law scaling of average properties with the
chain size. Local properties were on the scale of a bead. We specifically
studied
the inter-bead separations, angles, shape and positional dynamics. Global or
macroscopic properties were on the scale of the entire chain. In particular our
study probed the size and overall motion of vibrated rings. 

The size of various parts of the experimental apparatus limits our study
to chains of moderate bead numbers typically $N < 300$ for rings. Consequently
the rings are analogous to relatively short self-avoiding walks. Despite the
relatively
short chain lengths, we are still able to deduce reasonable scaling laws.
The mean global properties typically show power-law scaling over the modest
range in $N$ with cross-over effects at small $N$ and pronounced finite-size
effects at large $N$. Generally in between the extremes, power-law scaling
appears to provide an adequate description.

The system of vibrating granular chains presents a fertile ground for
experimentally studying nonequilibrium statistical mechanics. It is particularly
well-suited since it permits simultaneous measurements of micro- and
macro-scopic variables. Our future experiments will primarily concern this
aspect in the setting of multi-chain experiments. We expect to study groups of
rings and free-end chains and measure the average individual and collective
properties as they vary with density and conformation. The variation of the
average individual
or microscopic properties with the
macroscopic density will elucidate the statistical mechanics of this system.
Rings of different $N$ have different internal degrees of freedom. A tight ring
with $N=8$ is almost circular and has small
deviations from this configuration under vibrational excitation. Its space of
configurations
is small and we may think of it as having almost zero internal degrees of
freedom. Rings of larger $N$, however, have shapes that depart significantly
from
simple conformations and consequently a large configuration space and thus many
internal degrees of freedom. The richness bestowed by the ability to vary these
internal degrees of freedom will permit the detailed study of their bearing on
macroscopic properties of a collection of such rings. Other cooperative behavior
arising from rings of different sizes, a collection free-end chains and
self-interactions between different regions of very long chains afford abundant
possibilities
for studying nonequilibrium statistical physics.

We thank Mike Rivera and Brent Daniel for useful suggestions regarding the
experiment and Matt
Hastings. This research is supported by the U.S. DOE (W-7405-ENG-3

\end{document}